\documentclass[twocolumn,pre,aps,showpacs,amsmath,amssymb]{revtex4-1}
\usepackage{graphicx}
\usepackage{bm}
\usepackage{epstopdf}
\usepackage{braket}
\usepackage[brazilian]{babel}
\usepackage[dvipsnames]{xcolor}
\usepackage{ulem}

\begin{document}

    \title{Heat distribution in  open quantum systems with maximum entropy production}

\author{D. S. P. Salazar$^1$, A. M. S. Mac\^edo$^{2}$ and  G. L. Vasconcelos$^3$ }
\affiliation{$^1$Unidade de Educa\c{c}\~ao a Dist\^ancia e Tecnologia, Universidade Federal Rural de Pernambuco, 52171-900
Recife, PE, Brazil,}
\affiliation{$^2$ Laborat\'orio de F\'{\i}sica Te\'orica e Computacional, Departamento de F\'{\i}sica, Universidade Federal de Pernambuco 50670-901 Recife, Pernambuco, Brazil,}
\affiliation{$^3$ Departamento de F\'{\i}sica,
Universidade Federal do Paran\'a, 81531-990 Curitiba, Paran\'a, Brazil}

\begin{abstract}
We analyze the heat exchange distribution of quantum open systems undergoing a thermal relaxation that maximizes the entropy production.  
We show that the process implies a type of generalized law of cooling in terms of a time dependent effective temperature $T_t$. Using a two-point measurement scheme, we find an expression for the heat moment generating function that depends solely on the system's partition function and on the law of cooling. 
  Applications include the relaxation of free bosonic and fermionic modes, for which closed form expressions for the time-dependent heat distribution function are derived. Multiple free modes with arbitrary dispersion relations are also briefly discussed. In the semiclassical limit our formula agrees well with previous results of the literature for the heat distribution of an optically trapped nanoscopic particle far from equilibrium.
\end{abstract}

\maketitle
\textit{Introduction.}---In recent years significant results related to the nonequilibrium statistics of entropy production in open systems have been obtained \cite{Jar1997,Jar2004,Crooks1999,Seifert2008,Sekimoto2010}. A cornerstone of the field is the entropy fluctuation theorem (FT) which states that there is a special constraint in the asymmetry of the entropy production. In its integral form, the entropy FT reads $\langle e^{-\Delta_i S} \rangle = 1$, where $\Delta_i S$ is the  entropy produced in some nonequilibrium thermodynamic process  and the average is over all system's stochastic trajectories. From Jensen's innequality, the fluctuation theorem implies the Clausius inequality: $\Delta_i S \geq 0$. Important particular cases of the FT are the Jarzinski equality \cite{Jar1997}, the heat exchange fluctuation theorem \cite{Jar2004} and other less general forms \cite{Esposito2009,Cuetara2014}, thus making the integral FT (and its detailed-balance versions \cite{Crooks1999}) a central result in stochastic thermodynamics \cite{Seifert2012,Esposito2010,Esposito2012}. 

Nonequilibrium thermodynamics was also applied to open quantum systems, including the analysis of time dependent statistics of quantities such as heat and work \cite{Caldeira2014,Yu1994,Hanggi2005,Campisi2009,Seifert2012}, as well as quantum heat engines \cite{Scully2003} and refrigerators \cite{Dong2015,Karimi2016}. Some exact nonequilibrium results have already been obtained experimentally \cite{Zhang2015} and a few analogous phenomena have been reported in the  quantum information literature \cite{Parrondo2015,Horodecki2013}. The extension of the FTs to open quantum systems is particularly subtle because of the role played in the theory by the measurement scheme \cite{Suomela2014,Elouard2017,LlobletPRL2017}. For instance, there are different definitions of thermodynamic work depending on the methods used to account for the measurement effect, such as two point measurement (TPM) schemes  and quantum jump methods \cite{Allahverdyan2005,Allahverdyan2014,Solinas2015,Miller2017,Hekking2013,Liu2014,Suomela2016}, besides the fact that work is not a proper quantum observable \cite{TalknerPRE2017}. It was only very recently that a path-integral formulation of quantum work \cite{Funo2018} allowed its consistent definition in the presence of strong coupling.

There are also nonequilibrium situations---such as thermal relaxation processes---in which the relevant observable is the quantum heat \cite{Ronzani2018}. In those cases, one can unambiguously argue that the thermodynamic work performed by/over the system is zero \cite{Elouard2017,Gherardini2018,Yi2013,Funo2018}. In such situation, the heat $Q$ exchanged with the reservoir can be identified with the energy variation between two projective measurements. In other words, if $E_n$ (at $t=0$) and $E_m$ (at $t>0$) are the energies obtained in two consecutive measurements, then the heat {\it absorbed} by the system is $Q=E_m-E_n$. The quantum heat $Q$ is a stochastic quantity and  obtaining a closed form expression for its time dependent distribution, $P_t(Q)$, is not a trivial task. In fact, we are unaware of any previous exact result  for the time-dependent heat distribution in quantum open systems. Furthermore, the current literature does not even offer a general qualitative understanding of nonequilibrium quantum heat distributions since they may depend on system specific properties \cite{Elouard2017}.  Knowledge of the distribution $P_t(Q)$ is of both practical and conceptual importance in that it may help to address fundamental questions arising from nonequilibrium thermodynamics, as in the description of quantum engines \cite{Uzdin2015,Campisi2015}, single ion measurements \cite{Huber2008,Abah2012,Zhang2014}, and the estimation of the probability of an apparent violation of the second law of thermodynamics in small systems \cite{Brandao2015,Rivas2017}.

\begin{figure}[t]
\includegraphics[width=3.3 in]{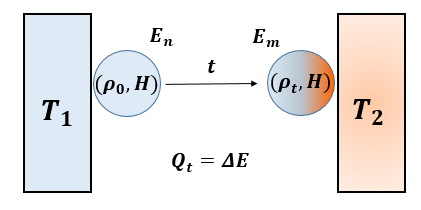}
\caption{(Color online) A system ($\rho_0,H$) is initially prepared in thermal equilibrium with the first (blue) reservoir at temperature $T_1$. At $t=0$, a projective energy measurement is performed, yielding the value $E_n$. Then, the system is placed in contact with a second (red) reservoir at temperature $T_2$. At time $t>0$, a second projective energy measurement is performed, yielding the value $E_m$. In this two-point measurement scheme, the heat exchanged between the system and the second reservoir is a time dependent stochastic quantity $Q_t=E_m-E_n$ described by a nonequilibrium probability distribution $P_t(Q)$.}
\label{fig1}
\end{figure}

In this letter, we compute the distribution $P_t(Q)$ for a wide class of thermalization processes satisfying a maximum entropy production principle (MEPP) \cite{Martyushev2006,Reina2015}. To be specific, we consider a two-point measurement (TPM) scheme (see Fig.~\ref{fig1}) where the system  is initially in thermal equilibrium with a reservoir of inverse temperature $\beta_1=1/T_1$. (Here  the Boltzmann constant is set to unity: $k_B=1$.) At  $t=0$ a measurement is made yielding, say, an  energy $E_n$, after which  the system is placed in  contact with a second reservoir with inverse temperature $\beta_2$. At some later time $t>0$ a second measurement is  performed yielding an energy $E_m$. Repeating this procedure several times allows one to construct the heat probability distribution $P_t(Q)$, where $Q=E_m-E_n$. First we show that under the assumption that  the net heat transfer $\langle Q\rangle_t$ is fixed the MEPP implies that  the system's density matrix $\rho_t$ remains thermal for $t\geq 0$, being characterized by a time-dependent effective  temperature $\beta(t)=\phi_t(\beta_1,\beta_2)$, for some function $\phi_t$ that describes the specific ``law of cooling'' of the system in question. Note in particular that $\phi_t$ must obey the following relations: i) $\phi_0(x,y)=x$; ii) $\phi_\infty(x,y)=y$; and iii) $\phi_t(x,x)=x$.

This general law of cooling has been found in several dynamics, such as in quasistatic processes \cite{Cugliandolo1993,Cugliandolo1997}, in Lindblad's dynamics of free bosonic \cite{Isar1994} and fermionic modes \cite{Earl2015}, in general time evolution of glassy systems \cite{Nieuwenhuizen1998}, in dynamical models for sheared foam \cite{Ono2002}, in the classical dynamics of optically trapped nanoparticles with experimental confirmation \cite{Gieseler2012,SalazarLira1,Gieseler2018}, and in the dynamics of some graphene models \cite{Sun2008}.

For such broad class of systems that evolve through thermal states, we show that the nonequilibrium quantum heat statistics is, quite surprisingly, entirely determined by the equilibrium partition function, $Z(\beta)=\sum_n e^{-\beta E_n}$. More precisely, we demonstrate that the time-dependent moment generating function (MGF), $M(s,t)=\langle e^{sQ}\rangle_t\equiv\int e^{sQ}P_t(Q)dQ$,  of the quantum heat $Q$ satisfies the remarkable identity:
\begin{equation}
\label{eq:M}
M(s,t)= \frac{Z(\beta_1+s)}{Z(\beta_1)}
\frac{Z(\phi_t(\beta_1+s,\beta_2)-s)}{Z(\phi_t(\beta_1+s,\beta_2))}.
\end{equation}
Two special cases of relation (\ref{eq:M}) are worth noting. First, setting $s=0$ yields $M(0,t)=1$, as required from the normalization condition: $\int_{-\infty}^\infty P_t(Q) dQ=1$. Secondly, for $s=\Delta \beta=\beta_2-\beta_1$ we recover the integral fluctuation theorem \cite{Jar2004}: 
\begin{align}
M(\Delta \beta,t)=  \langle  e^{ \Delta \beta Q} \rangle = 1,
    \label{eq:IXFT}
\end{align}
which follows immediately from the fact that $\phi_t(\beta_2,\beta_2)=\beta_2$ for all $t$. 
Furthermore, for  systems that satisfy detailed balance (DB), meaning that ${P_t(n\rightarrow m)}=e^{-\beta_2(E_m-E_n)} {P_t(m\rightarrow n)}$,
where $P_t(n\rightarrow m)$ denotes the transition probability from state $|n\rangle$ to $|m\rangle$,
one can show that the MGF possesses the following symmetry:
\begin{align}
    M(\Delta \beta -s,t)=M(s,t),
    \label{eq:sym}
\end{align}
which is a direct manifestation of the detailed fluctuation theorem (DFT) \cite{Salazar2017,iranianos2018}: ${P_t(Q)}/{P_t(-Q)}=e^{-\Delta \beta Q}$. Note, however, that the DFT, as expressed in (\ref{eq:sym}), is a general property of systems obeying DB, whereas  Eq.~(\ref{eq:M}) is a stronger result that relates non-equilibrium fluctuations with the equilibrium distribution for systems that obey a thermalization dynamics (not necessarily satisfying DB).
Relation (\ref{eq:M}) can thus be seen as a generalized fluctuation theorem that shows that
the time-dependent heat distribution is fully encoded in the equilibrium partition function and in the underlying law of cooling, which accounts for the weak coupling with the reservoir. This result has important practical consequences, as it allows us to compute the non-equilibrium heat distribution $P_t(Q)$  for several systems of interest, as  will be shown later.

\textit{Maximum entropy production.}---We consider open quantum systems \cite{Rivas2012} whose time evolution is described by a  dynamical map,  $\rho_t=\Phi_t(\rho_0)$, with the semigroup property: $\Phi_t(\Phi_s(\rho))=\Phi_{t+s}(\rho)$, for $s,t\geq 0$.
This implies in practice that for $t>0$ the system is weakly coupled to the heat bath and that the  evolution is memoryless (i.e., Markovian). The dynamics acts on the system starting at a thermal state with an inverse temperature $\beta_1$: $\rho_0=\rho^{\beta_1}\equiv e^{-\beta_1 H}/Z(\beta_1)=\left[Z(\beta_1)\right]^{-1}\sum_n e^{-\beta_1 E_n} |n\rangle \langle n| $. We analyze relaxation processes $\Phi_t$ with target state $\rho^{\beta_2}$, representing the second heat bath at inverse temperature $\beta_2$, so that $\Phi_t(\rho^{\beta_2})=\rho^{\beta_2}$ for all $t$. We shall consider that the dynamics $\Phi_t$ is such that the entropy production in the interval $[0,t]$ is maximal, given that the net heat flux, $\langle Q \rangle_t$, is fixed. This assumption is consistent with the existence of fundamental geometric bounds on the entropy production of open quantum systems, which has a been the central result of a recent work \cite{Mancino2018}. We show below that under these assumptions the thermal relaxation $\Phi_t$ satisfies a generic ``law of cooling'' of the type
\begin{equation}
\label{cooling}
\Phi_t(\rho^{\beta_1})=
\rho^{\phi_t(\beta_1,\beta_2)},
\end{equation}
which maps an initial thermal state at inverse temperature $\beta_1$ onto a thermal state with an effective (time-dependent) inverse temperature $\beta(t)=\phi_t(\beta_1,\beta_2)$.

To establish (\ref{cooling}), we first recall that in the TPM scheme the system is subjected to a projective measurement at $t=0$ and subsequently placed in contact with another heat bath at temperature $T_2$, whose dissipative dynamics is represented by the map $\Phi_t$. At time $t>0$, the non-equilibrium density matrix $\rho_t=\Phi_t(\rho_0)$ has an associated von Neumann's entropy variation $\Delta S=S[\rho_t]-S[\rho_0]$, where $S[\rho]=-{\rm tr}(\rho \log \rho)$. One can write the entropy variation as $\Delta S = \Delta_i S + \Delta_e S$, where $\Delta_i S$ is the entropy production (i.e., the irreversible component of the entropy change) and $\Delta_e S=\beta_2 \langle Q \rangle_t$  is the entropy exchanged with the environment (reservoir $2$), corresponding to the reversible contribution to the entropy variation, with  $\langle Q \rangle_t=\langle H \rangle_t -\langle H \rangle_0={\rm tr}(\rho_t H)-{\rm tr}(\rho_0 H)$ being the net heat absorbed by the system. According to   Clausius inequality,  the entropy production is positive: $\Delta_i S\geq 0$.
(We remark parenthetically that sharper lower and upper bounds for the irreversible entropy production have been recently   established for open quantum systems  \cite{lutz2010, Mancino2018}.) Here we shall require $\Delta_i S$ to be maximal for a constant heat exchange $\langle Q \rangle_t$. Thus,  maximizing $\Delta_i S$ is equivalent to maximizing $\Delta S$. This optimization problem may  be written in the usual form: $\delta (S_t -a_t \langle H \rangle_t+b_t\langle 1 \rangle)=0$,
where $\delta F(p)$ is understood as the functional derivative of $F(p)$ with respect to the distribution $p$, and the Lagrangian multipliers $a_t$ and $b_t$ are needed to account for the constraint on the heat flux $\langle Q \rangle_t$ (and hence on $\langle H \rangle_t$ since $\langle H \rangle_0$ is fixed by the initial state) and the normalization condition, respectively. Solving the optimization with these constraints results in a time-dependent thermal density matrix $\rho_t=e^{-a_t H}/Z(a_t)$, where $Z(a_t)={\rm tr}[e^{-a_t H}]$, with $a_0=\beta_1$ and $a_\infty=\beta_2$. 
From the constraint on $\langle H \rangle_t$, one can define $a_t$ for any $t>0$ from the formula $\langle H \rangle_t=-\partial \ln Z(a_t)/\partial a_t$. Solving for $a_t$  yields the law of cooling $a_t=\phi_t(\beta_1,\beta_2)$.
Note that this is equivalent to the property defined in (\ref{cooling}), which states that the dynamical map $\Phi_t$ evolves initial thermal states onto thermal states  for all $t>0$.

It is perhaps worth pointing out that the above argument remains valid if instead of the von Neumann entropy one uses  the Wigner entropy production, which has the advantage  that the  entropy flux stays finite for $T\to 0$ \cite{Paternostro2017}. The only difference  is that in the case of the Wigner entropy one has $\Delta_e S = f(\beta_2) \langle Q \rangle$, where  the function $f(\beta_2)$ depends on the partition function of the system \cite{Paternostro2017}. Hence a MEPP in terms of the Wigner entropy also leads to a law of cooling as in (\ref{cooling}). In fact, it has been recently shown that thermalization is a rather general mechanism in quantum systems under a measurement process \cite{japoneses2018}, and so the existence of an effective temperature applies to a  broad class of open systems.

\textit{Heat distribution.}---Now that the dynamical map $\Phi_t$ has been shown to imply a law of cooling as in (\ref{cooling}), we prove that the heat distribution obeys the  fluctuation relation given in (\ref{eq:M}). We recall that in the TPM scheme, the system starts at thermal equilibrium with the first reservoir at temperature $T_1$, and  at $t=0$ an energy measurement is performed yielding the value $E_n$ with probability $p_n={e^{-\beta_1 E_n}}/Z(\beta_1)$, and thus projecting the system onto the energy eigenstate $|n\rangle \langle n |$. Subsequently, the system is placed in thermal contact with a second reservoir at temperature $T_2$, represented by the map $\Phi_t$ with the cooling property (\ref{cooling}). A second energy measurement is then performed on the system at some time $t>0$, now with the time propagated density matrix $\Phi_t(|n\rangle \langle n |)$, yielding the value $E_m$ and projecting the system onto the energy eigenstate $|m\rangle \langle m|$. The moment generating function of the exchanged heat $Q=E_m-E_n$ is defined as
\begin{equation}\label{mgf2}
\langle e^{sQ} \rangle = \sum_{n,m}e^{s(E_m-E_n)}p_n 
\langle m | 
\Phi_t\Big(|n\rangle \langle n|\Big)
|m\rangle.
\end{equation}
Using the linearity of $\Phi_t$ and combining the terms in $p_n$ and $e^{-sE_n}$, we can  rewrite the sum over $n$ in (\ref{mgf2}) in terms of a new thermal state that depends on the real parameter $s$ (provided $s+\beta_1\geq 0$), thus obtaining
\begin{equation}
\langle e^{sQ} \rangle = 
\frac{Z(\beta_1+s)}{Z(\beta_1)}
\sum_{m}e^{s E_m} 
\langle m | 
\Phi_t\Big( \rho^{\beta_1+s}\Big)
|m\rangle;
\label{eq:esQ}
\end{equation}
see details in the Supplemental Material.
Finally, we apply  property (\ref{cooling}) to write $\Phi_t(\rho^{\beta_1+s})=\rho^{\phi_t(\beta_1+s,\beta_2)}=e^{-\phi_t(\beta_1+s,\beta_2)H}/Z(\phi_t(\beta_1+s,\beta_2))$, which inserted into (\ref{eq:esQ}) and summing over $m$ results in Eq.~(\ref{eq:M}). Next, we  shall make use of the MGF (\ref{eq:M}) to compute explicitly the heat distribution $P_t(Q)$ for a variety of systems. 

\textit{Bosonic Modes.}---Here we apply (\ref{eq:M}) to a system composed of a single bosonic mode coupled to a thermal bath. The system is described by the Hamiltonian of the harmonic oscillator $H=\hbar\omega (a^\dagger a+1/2)$ and its partition function is $Z(\beta)=(1/2){\rm csch}(\beta \hbar \omega/2)$. In this representation, the system satisfies a Lindblad equation  \cite{Paternostro2017}
\begin{equation}
\label{Lind}
\partial_t \rho = \frac{-i}{\hbar}[H,\rho] + D_i(\rho)
\end{equation}
with the dissipator
\begin{equation}
D_i(\rho)=\gamma(\overline{n}_i+1)[a\rho a^\dagger - \frac{1}{2}\{a^\dagger a, \rho\}]+\gamma \overline{n}_i[a^\dagger \rho a -\frac{1}{2}\{a^\dagger a , \rho\}],
\label{Dissipator}
\end{equation}
and an average number of excitations, $\overline{n}_i=(\exp(\hbar \omega/k_BT_i)-1)^{-1}$, $i=1,2$, that depends on the temperature of the reservoir coupled to the system. One can verify using the operator formalism \cite{Louisell1965,Isar1994} that the dynamics (\ref{Lind}) does indeed satisfy a law of cooling of the type shown in (\ref{cooling}). 
This  result can be obtained more directly using the Wigner function representation and its associated stochastic parametrization. More specifically, one can show (see Sup.~Mat.) that the dynamics given by (\ref{Lind}) and (\ref{Dissipator}) propagates any thermal distribution with inverse temperature $\beta_1$ to another thermal distribution with an effective time-dependent temperature defined by $\overline{n}_t=(\overline{n}_1-\overline{n}_2)e^{-\gamma t}+\overline{n}_2$, where $\overline{n}_t=(\exp(\hbar \omega/k_BT_t)-1)^{-1}$,  which can be solved to yield $\beta_t\equiv1/(k_B T_t)= \phi_t(\beta_1,\beta_2)$.

\begin{figure}[t]
\includegraphics[width=3.3 in]{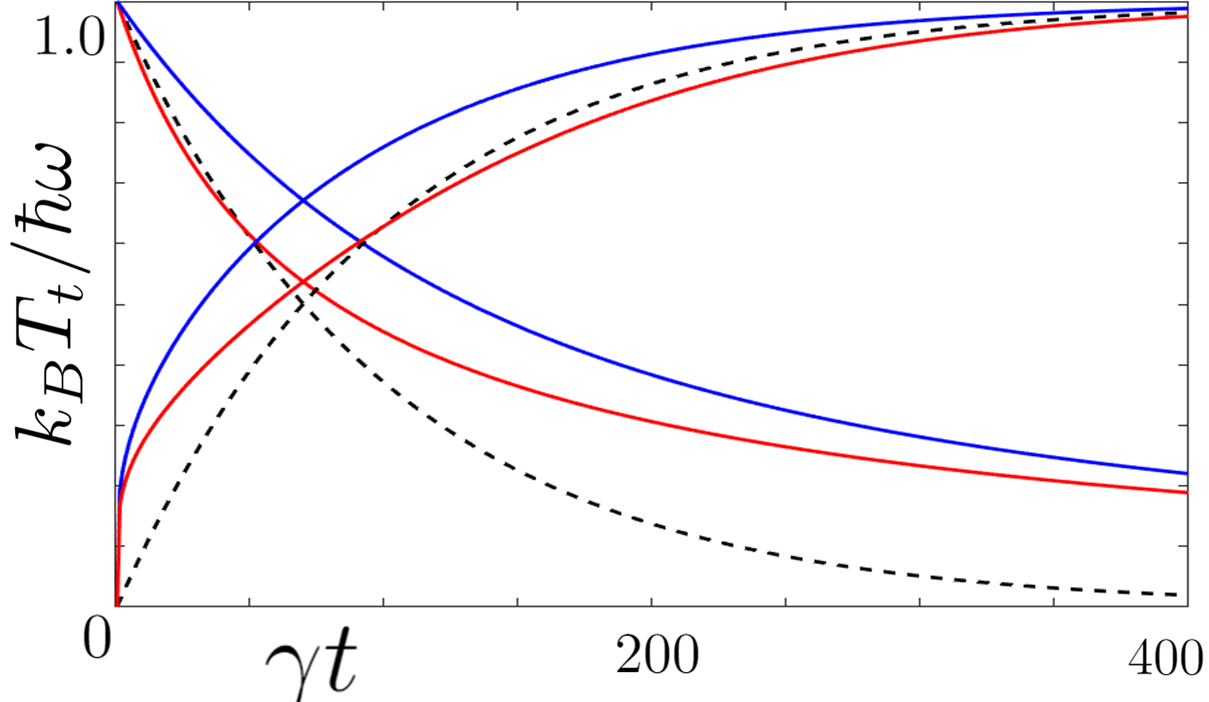}
\caption{(Color online) Different laws of cooling  (effective temperatures) of the bosonic mode (solid blue), fermionic mode (solid red) and the classical Newton's law of cooling (dashed) as a function of time. The descending lines represent cooling from $k_BT_1/\hbar \omega=1$ to $k_B T_2 /\hbar \omega=10^{-3}$. The rising lines represent the heating process from $k_BT_1/\hbar \omega = 10^{-3}$ and $k_BT_2/\hbar \omega = 1$. Note that the bosonic line (blue) decays slower in the cooling process and rises faster in heating when compared to its semiclassical limit (dashed). The fermionic starts by heating faster, but then becomes slower than the semiclassical limit. In the cooling process, the fermionic mode starts close to the classical case, but eventually slows down.}
\label{fig2}
\end{figure}

It is worth mentioning that in the semiclassical limit, where $\hbar \omega (\overline{n}_i+1/2)\rightarrow k_B T_i$,  one recovers the Newton's law of cooling $T_t=T_2+(T_1-T_2)e^{-\gamma t}$.
Note that although the effective temperature $T_t$ evolves from $T_1$ (at $t=0$) and reaches $T_2$ for $t\to\infty$ in both the quantum and the classical cases, the transient behaviors differ considerably, as depicted in Fig.2.

Having found that the system (\ref{Lind}) satisfies a law of cooling as in (\ref{cooling}), a remarkably elegant expression for the heat MGF $M(s,t)$ can now be obtained by using Eq. (\ref{eq:M}) with $Z(\beta)=(1/2){\rm csch}(\beta \hbar \omega/2)$. We find  
\begin{equation}
\label{mgf}
M(s,t) =
\left[1-\frac{\sinh(\frac{\hbar}{2}\omega s)\sinh(\frac{\hbar}{2}\omega (s-\Delta \beta))}{\sinh(\frac{\hbar}{2}\omega \beta_1)\sinh(\frac{\hbar}{2}\omega\beta_2)}(1-e^{-\gamma t})\right]^{-1},
\end{equation}
where $\Delta \beta = \beta_2-\beta_1$. This equation  reproduces, after some algebra, a recent result reported in \cite{Denzler2018}. As a check, note that $M(0,t)=1$, whereas for $s=\Delta \beta$ we indeed obtain (\ref{eq:IXFT}). Furthermore, Eq.~(\ref{mgf}) displays the symmetry  (\ref{eq:sym}), as expected, since the system is known to satisfy detailed balance. 

Remarkably, we are also able to find (see Sup.~Mat.) the nonequilibrium probability distribution of the exchanged heat, $Q=E_m-E_n=\hbar \omega (m-n)=\hbar \omega k$, a task that was deemed not possible in Ref.~\cite{Denzler2018}. We obtain

\begin{equation}
\label{pmf}
P_t(Q=\hbar \omega k) = \frac{2\nu_t\mu_t}{\mu_t^2-1}\exp\big(-|k|\ln\mu_t- k\frac{\hbar \omega\Delta \beta}{2}\big),
\end{equation}
for any integer $k=\{...,-1,0,+1,...\}$, with time dependent parameters $(\nu_t,\mu_t)$ given by $\nu_t=2\sinh(\hbar\omega\beta_1/2)\sinh(\hbar\omega\beta_2/2)/(1-e^{-\gamma t})$, $\mu_t=\Omega_t+\sqrt{\Omega_t^2-1}$, and $\Omega_t=\nu_t+\cosh(\hbar \omega \Delta \beta/2)$. Note in particular that the DFT holds: $P_t(Q=\hbar\omega k)/P_t(Q=-\hbar \omega k) = e^{-Q\Delta \beta}$.
Notably, we may also use (\ref{pmf}) to find the probability of a heat flow from lower to higher temperature. To see this, suppose that $T_2>T_1$. In this case, one expects a positive heat ($Q>0$) absorbed by the system. However, since $Q$ is a random variable, there is a probability of a reverse heat flow ($Q\leq 0$) given by
\begin{equation}
\label{reverseQ}
P_t(Q\leq 0)=\frac{2\nu_t\mu_t}{\mu_t^2-1}\frac{1}{(1-\mu_t^{-1}e^{\hbar \omega\Delta \beta/2})}.
\end{equation}
This apparent violation of the second law of thermodynamics is indeed observed in small systems \cite{Evans2002}.
Before leaving this section, we remark that a perturbation in the Hamiltonian which is linear in $a$ and $a^\dagger$, such as in the  case of a single mode cavity pumped by a radiation field, does not change the heat MGF (\ref{mgf}) since the pump only shifts the spectrum $E_n$ by a constant, which keeps the energy variations ($\Delta E = E_m-E_n$) invariant. Thus, relation (\ref{mgf}) can in principle be tested in optical cavities far from equilibrium.

\textit{Underdamped classical oscillators.}---The semiclassical limit of Eq.~(\ref{mgf}) is obtained by taking $\hbar\omega\beta_i \rightarrow 0$ and $\hbar\omega s \rightarrow 0$. We find
\begin{equation}
\label{mgfClassic}
M(s,t)=
\big(1-\frac{s(s-\Delta \beta)}{\beta_1\beta_2}(1-e^{-\gamma t})\big)^{-1}.
\end{equation}
Note in particular that in the case of $f/2$ independent  harmonic oscillators we have $M_{f/2}(s,t)=\left[M(s,t)\right]^{f/2}$, which combined with (\ref{mgfClassic}) reproduces the result for the heat distribution of $f/2$ classical nanoscopic particles optically trapped in the highly underdamped limit \cite{Gieseler2012,SalazarLira1,Gieseler2018}.

\textit{Fermionic mode.}---Consider a free fermionic mode with Hamiltonian $H=\hbar \omega a^\dagger a$ and usual anticommutating relation $\{a^\dagger, a\}=1$. The system has two eigenstates $\{|0\rangle, |1\rangle\}$. Given a initial state $v=(1-\overline{n},\overline{n})$, suppose its dynamics is Markovian and satisfies detailed balance. Then, starting at an initial temperature $T_1$, the evolution of the state is given by the same law of cooling found in the bosonic case, namely $\overline{n}_t=\overline{n}_2+(\overline{n}_1-\overline{n}_2)e^{-\gamma t}$, but where now ${n}_i$ is the fermionic occupation number: $\overline{n}_i=(\exp(\beta_i \hbar \omega)+1)^{-1}$, $i=1,2$, for some damping constant $\gamma>0$, obtained from the unique free parameter in the dynamics (see Sup. Mat.). The fermionic law of cooling is also depicted in Fig.~\ref{fig2}. In this case, the heat exchange MGF reads
\begin{equation}
\label{fermionmgf}
M(s,t) = 1+\frac{\sinh(\hbar\omega s/2)\sinh(\hbar\omega(s-\Delta \beta)/2)}{\cosh(\hbar\omega\beta_1/2)\cosh(\hbar\omega\beta_2/2)}(1-e^{-\gamma t}).
\end{equation}
Note the striking similarities between this result and that for the bosonic case shown in (\ref{mgf}). In particular, Eq~(\ref{fermionmgf}) has  the expected behavior for  $s=0$ and $s=\Delta \beta$, and satisfies (\ref{eq:sym}) as well. The heat probability distribution is found straightforwardly from (\ref{fermionmgf}) to be
\begin{equation}
\label{fermionpmf}
P_t(Q=\pm \hbar \omega)=
\frac{(1-e^{-\gamma t})e^{\mp \Delta \beta/2}}
{4\cosh(\hbar \omega \beta_1/2)\cosh(\hbar \omega \beta_2/2)},
\end{equation}
with $P_t(Q=0)=1-(P_t(\hbar\omega)+P_t(-\hbar \omega))$.

\textit{Conclusions and Perspectives.}---We showed that a thermal relaxation process that maximizes the entropy production (for fixed heat exchange) satisfies a law of cooling of the form shown in (\ref{cooling}). Then, we proved that any system relaxing in this way has a heat moment generating function given by Eq.~(\ref{eq:M}), which depends only on the equilibrium partition function $Z(\beta)$  and the cooling function $\phi_t(\beta_1,\beta_2)$.  Finally, we used this general result to find the heat distribution of a number of experimentally relevant systems, such as a single bosonic mode, the semiclassic limit of optically trapped nanoparticles, and a single fermionic mode. We emphasize that the maximum entropy production principle may also be useful in deriving approximate cooling laws for a large class of systems with tight bounds on the entropy production \cite{Mancino2018}, where a nonequilibrium effective temperature can be defined.
Further applications may include systems coupled to particle reservoirs with different electrochemical potentials $(\mu_1,\mu_2)$. This situation is remarkably similar to the bosonic system treated in the paper, since the energies $E_n \propto n$ yields the same spectrum ($E_{n+1}-E_n=cte$) in both cases, and will be discussed in future studies. As a closing remark, we point out that the framework derived here can be easily generalized to include the study of quantum heat statistics of system with multiple independent modes, such as Bose-Einstein condensates and the relaxation dynamics of spin chains. This interesting perspective will be developed in future work.



{}
\end{document}